\documentclass[10pt,byrevtex,twocolumn,noshowpacs,prb]{revtex4}
\usepackage{amssymb}
\usepackage{graphics}
\usepackage{epsfig}
\usepackage{graphicx}
\usepackage{setspace}

\begin{document}

\title{Giant Carrier Mobility in Single Crystals of FeSb$_{2}$}
\author{Rongwei Hu$^{1,2}$, V. F. Mitrovi\'{c}$^{2}$, and C. Petrovic$^{1}$}
\affiliation{$^{1}$Condensed Matter Physics, Brookhaven National Laboratory, Upton, New
York 11973, USA\\
$^{2}$Physics Department, Brown University, Providence RI 02912, USA}
\date{\today }

\begin{abstract}
We report the giant carrier mobility in single crystals of FeSb$_{2}$.
Nonlinear field dependence of Hall resistivity is well described with the
two-carrier model. Maximum mobility values in high mobility band reach $\sim
10^{5}$ cm$^{2}$/Vs at 8 K, and are $\sim 10^{2}$ cm$^{2}$/Vs at the room
temperature. Our results point to a class of materials with promising
potential for applications in solid state electronics.
\end{abstract}

\maketitle

FeSb$_{2}$ is a narrow band nearly magnetic or Kondo semiconductor with 3d
ions.\cite{Petrovic1} It crystallizes in \textit{Pnnm} orthorhombic
structure and shows pronounced anisotropy in transport properties. Along the
high conductivity axis, metal-insulator transition onsets in the vicinity of
T$^{\ast }=40$ K, while the electronic transport along two other axes is
semiconducting for temperatures $(T)$ up to 350 K.\cite{Perucchi}$^{,}$\cite%
{Petrovic2} Moreover, the doping strongly affects the properties of FeSb$%
_{2} $.\cite{RongweiDop1}$^{,}$\cite{RongweiDop2} For example, colossal
magnetoresistance (CMR) of the same order of magnitude as in manganite
oxides and multiband transport properties are observed in Fe$_{1-x}$Co$_{x}$%
Sb$_{2}$ (x = 0 - 0.5).\cite{Salamon}$^{,}$\cite{Rongwei} In this work, we
report high Hall mobility in FeSb$_{2}$ that reaches 77 352 $\pm $ 547 cm$%
^{2}/$Vs, the value comparable to that of the Si/SiGe heterostructures.\cite%
{Ismail}

\begin{figure}[tbp]
\centerline{\includegraphics[height=4.6in]{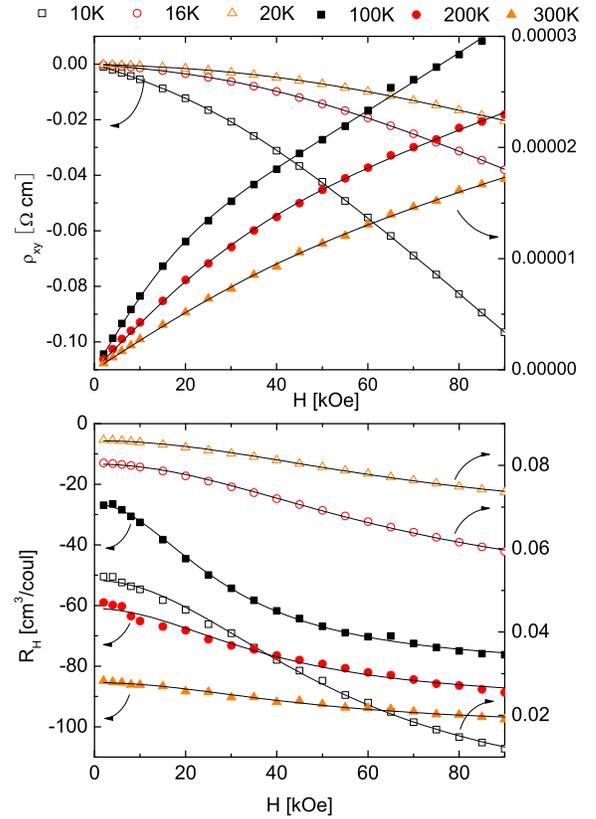}} 
\vspace*{-0.2cm}
\caption{a) Hall resistivity $\rho _{xy}$ as a function of
the applied field for selected temperatures. The solid lines are fits to the
two-carrier model, \textit{i.e.} H $\times $ Eq. 1. b) The Hall coefficient
as a function of the applied field.}
\end{figure}

Single crystals of FeSb$_{2}$ were grown from Sb flux.\cite{Petrovic2}$^{,}$%
\cite{Rongwei} Crystals were polished into rectangular bars and Pt wires
were attached using Epotek H20E silver epoxy. The electrical and Hall
resistivity were measured in Quantum Design PPMS instrument using four wire
and five wire configuration for AC measurements, respectively. The current
was applied along high conductivity $c$-axis. The magnetic field was applied
successively in $[1\overline{1}0]$ and $[\overline{1}10]$ directions and the
transverse voltage was picked up in the orthogonal [110] direction. One half
of the transverse voltage difference for field in two directions was taken
as the Hall voltage. The magnetization ($M$) was measured in Quantum Design
MPMS in the applied magnetic field of H = 1000 Oe along three principal
crystallographic axes. Polycrystalline magnetization was calculated as the
average of $M$ measured along the principal axes.

The magnetic field dependence of the Hall resistivity ($\rho _{xy}$) is
shown in Fig. 1(a). A nonlinear field dependence of $\rho _{xy}$ for $H$ up
to 90 kOe is evident. This result commonly indicates an anomalous Hall
effect. As in the case of Fe$_{1-x}$Co$_{x}$Sb$_{2}$, we discard this
possibility by comparing experimental curves with the those obtained by
fitting $\rho _{xy}$ data to $\rho _{xy}(H)=R_{0}H+R_{S}M(H)$, where R$_{0}$
and R$_{S}$ are the normal and spontaneous Hall constant and $M$ is the
experimentally obtained sample magnetization.\cite{Rongwei} Nevertheless,
the nonlinear magnetic field dependence of $\rho _{xy}$ is often observed in
multi-carrier materials, such as semiconducting heterostruture devices, InP
pHEMT device, and MnAs.\cite{Della}$^{,}$\cite{Berry} Therefore, our data
implies that the nontrivial $H$ dependence of $\rho _{xy}$ originates from
the presence of different carriers. We now proceed to the analysis of the
data in the multi-carrier scenario.

\begin{figure}[tbp]
\centerline{\includegraphics[height=3in]{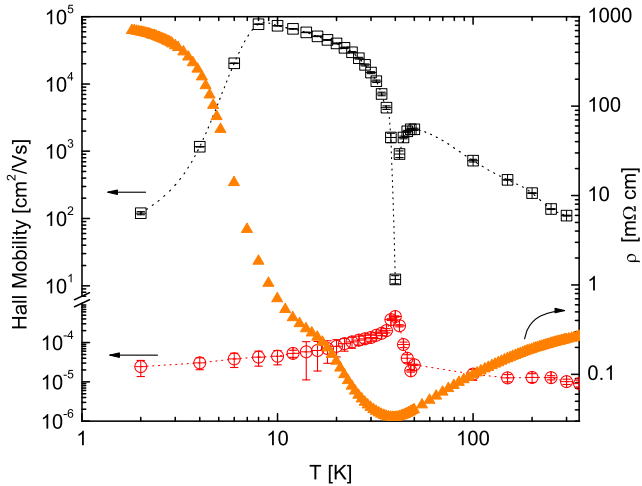}} 
\vspace*{-0.2cm}
\caption{The Hall mobility for two types of carriers denoted by the open
symbols. The nature of the carrier is denoted by the center +/- sign of the
data point,\textit{i.e.} + for holes and - for electrons. The resistivity
(filled symbols) is denoted on the right vertical axis.}
\end{figure}

In a two-carrier model, the Hall coefficient can be expressed in the
following form 
\begin{eqnarray}
\rho _{xy}/H &=&R_{H}=\rho _{0}\frac{\alpha _{2}+\beta _{2}H^{2}}{1+\beta
_{3}H^{2}}  \label{ResH} \\
\alpha _{2} &=&f_{1}\mu _{1}+f_{2}\mu _{2}  \nonumber \\
\beta _{2} &=&(f_{1}\mu _{2}+f_{2}\mu _{1})\mu _{1}\mu _{2}  \nonumber \\
\beta _{3} &=&(f_{1}\mu _{2}+f_{2}\mu _{1})^{2}.  \nonumber
\end{eqnarray}%
Here $\rho _{0}$ is the zero field resistivity, $\mu _{i}$ the mobility of i$%
^{th}$ carrier, and $f_{i}=|n_{i}\mu _{i}|/\sum |n_{i}\mu _{i}|$ the $f$
factor.\cite{JSKim} We point out that the term \textquotedblleft carrier" is
defined as a set of carriers having the same mobility. Therefore, one
carrier is associated with only one energy and/or one degenerate energy
level. It is different from the conventional electron or hole carriers, that
may relate to a continuous energy band. An excellent agreement of the R$_{H}$
field dependence with the two-carrier model Eq. 1 is illustrated in Fig.
1(b). Thus, the mobility $(\mu )$ and concentration $(n)$ of individual
carriers can be extracted.

The temperature dependence of the mobility is shown in Fig. 2. It was
generated from a large number of Hall resistivity isotherms. At room
temperature two types of charge carriers, high mobility ($\mu _{H}\sim
10^{2} $ cm$^{2}$/Vs) and low mobility ($\mu _{L}\sim 10^{-5}$ cm$^{2}$/Vs)
ones, are present. The low mobility carriers display a relatively weak
temperature dependence. The magnitude of $\mu _{L}$ slightly increases only
in the vicinity of T$^{\ast }=$ 40 K, temperature of metal to insulator
crossover. On the other hand, the high mobility component, $\mu _{H}$,
displays a significant temperature dependence. As $T$ decreases from 300 K,
its value increases from $\mu _{H}$(300K)=110 $\pm $ 2cm$^{2}$/Vs to reach a
remarkably high maximum value of $\mu _{H}$(8K)=77352 $\pm $ 547 cm$^{2}$/Vs
at T = 8 K. Below T = 8 K, $\mu _{H}$ decreases and at T = 1.8 K, lowest $T$
investigated, it reaches the value comparable to the room temperature one.
In contrast to the low mobility carriers, the sign of $\mu _{H}$ changes
from negative to positive in the vicinity of T$^{\ast }$. That is to say the
conduction changes from the electron to hole-like. Mobility and carrier
concentration changes at T$^{\ast }$ are rather sharp in comparison with
crossover -- like feature in resistivity and magnetic susceptibility.
Simultaneous increase of mobility in low mobility band and decrease of
mobility in high mobility band as well as changes from hole to electron
conduction in large mobility band around T$^{\ast }$ could contribute to
somewhat broad resistivity transition. Large increase in thermally induced
paramagnetic moment occurs above T$^{\ast }$ possibly due to indirect energy
gap.

\begin{figure}[tbp]
\centerline{\includegraphics[height=3in]{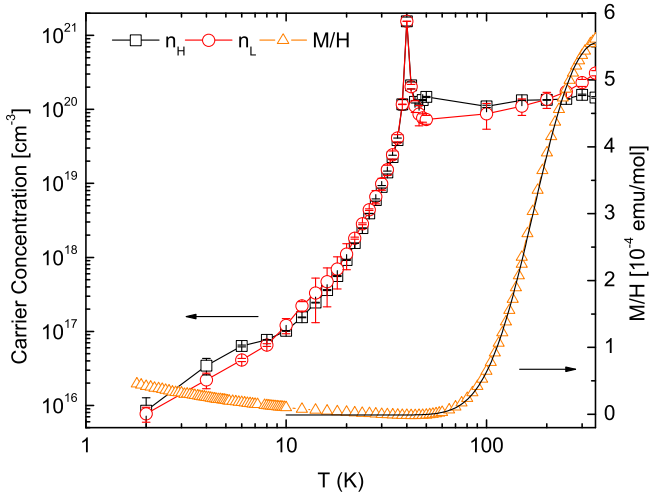}} 
\vspace*{-0.2cm}
\caption{Carrier concentrations of high and low mobility carriers as a
function of temperature. The polycrystalline average of the magnetic
susceptibility is denoted on the right axis. The solid line is the fit to
narrow band small gap model, that yields W = 278 K and $E_{g}$ = 2$\Delta $
= 856 K.}
\end{figure}

The temperature dependence of the carrier concentrations corresponding to $%
\mu _{L}$ and $\mu _{H}$ is shown in Fig. 3. Large increase of magnetic
susceptibility above T$^{\ast }$ coincides with small changes in carrier
concentration above 100 K. There is little change in M/H below T$^{\ast }$.
It appears therefore that significant reduction in carrier concentration
below T$^{\ast }$ predominantly influences conductivity (Fig. 2). This may
indicate larger (direct) energy gap in M/H and smaller (indirect) energy gap
for electronic transport, as observed in FeSi.\cite{Paschen} In the metallic
region of conductivity, from room temperature down to T$^{\ast }$,
concentrations of both low, $n_{L}$ (electron-like above T$^{\ast }$), and
high, $n_{H}$ (hole-like above T$^{\ast }$), mobility carrier components
slowly decrease. Below T$^{\ast }$, both $n_{L}$ and $n_{H}$ decrease by
approximately five orders of magnitude. The peak in the $n_{L}$ coincides
with the sign change of mobility at the T$^{\ast }$. This suggests that the T%
$^{\ast }$ is related to the intrinsic reconstruction of the electronic
properties of FeSb$_{2}$ at the metal to insulator crossover. This
observation, together with the fact that the concentration of low mobility
electron carriers increases by one order of magnitude at T$^{\ast }$, may
imply that above T$^{\ast }$ the Fermi level is situated in a band with a
larger density of states. This interpretation of FeSb$_{2}$ is consistent
with the narrow band small gap Kondo description, based on a model of two
narrow bands at the density of states of width W separated by Eg=2$\Delta $.%
\cite{Jaccarino}$^{,}$\cite{Mandrus} This model is supported by the LDA+U
band structure calculation of FeSb$_{2\text{.}}$\cite{Lukoyanov} It appears
though that metal insulator crossover at T$^{\ast }$ has large effect on the
high mobility carrier system. As the sample is cooled down to $T=1.8\ K$,
the electrical resistivity increases more than four orders of magnitude from
its lowest value at T$^{\ast }$. The change of slope of $\mu _{H}(T)$ at $%
T=20$~K, as well as its pronounced decrease below $T=8$~K, is reflected in
corresponding changes in $\rho (T)$ below T$^{\ast }$. This may imply
successive destruction of several quasi - one dimensional pieces of the
Fermi surface below T$^{\ast }$, as reported, for example, in $\eta $-Mo$%
_{4} $O$_{11}$.\cite{Hill} The carrier concentrations and their temperature
dependencies are almost identical whereas mobilities differ by many orders
of magnitude. In the standard paradigm of Kondo Insulators\cite{Aeppli}
hybridization involves one flat band and one dispersive conduction electron
band. As opposed to FeSi, hybridization with Sb-p orbitals in FeSb$_{2}$
could create multiple sheets at the Fermi surface with large mobility
difference.

Conduction properties of marcasite antimonides are influenced by distortion
of edge-sharing Fe-Sb octahedra. The Fe-Sb-Fe bond angle governs the d$_{xy}$
orbitals' overlap, extending along the octahedral edges, and consequently
magnetic and electronic properties of FeSb$_{2}$.\cite{Petrovic1}$^{,}$\cite%
{Goodenough} Within the Kondo insulator framework, our results imply that
the hybridization in this multicarrier system is rather heterogenous and may
involve only one electronic subsystem, whereas the other is characterized by
much lower effective mass and consequently higher mobility. Orbital -
selective probes and high resolution local structural probes would be very
useful to shed more light to this problem. Synthesis of the iron
diantimonide thin films may enable fabrication of a variety of devices, that
exploit high mobility carrier channels. Moreover, due to low resistivity and
high mobility of carriers this material may be of interest in photovoltaics
industry, providing that the energy gap proves to be tunable.

In conclusion, we have demonstrated giant Hall mobility in correlated
electron or "Kondo" semiconductor FeSb$_{2}$. The mobility is remarkably
large, comparable to the high mobility heterostructures and semiconductor
devices. The observed magnetic field dependence of Hall coefficient can be
described by a two-carrier model. The presence of the metal - insulator
crossover temperature T$^{\ast }$ and quenching of temperature induced
paramagnetic moment is intimately connected with changes in Hall mobility
and carrier concentration of individual carrier systems. Our results point
to a class of materials showing promise for application in high-speed
electronic devices.

This work was carried out at the Brookhaven National Laboratory which is
operated for the U.S. Department of Energy by Brookhaven Science Associates
(DE-Ac02-98CH10886).

\end{document}